\documentclass[aps,twocolumn,amsmath,amssymb,superscriptaddress]{revtex4-2}

\usepackage{graphicx,relsize,epstopdf,color,mathtools,bm,newtxtext,newtxmath,braket,rotating,dsfont}

\usepackage{float}
\usepackage{flafter}
\usepackage{placeins}
\usepackage{soul}
\usepackage{cancel}
\usepackage[normalem]{ulem}

\usepackage[hyphenbreaks]{breakurl}
\usepackage[colorlinks=true,linkcolor=blue,citecolor=blue,urlcolor =blue]{hyperref}
\usepackage{pdfcomment}

\newcommand{\RB}{\mathrm{RB}}

\newcommand{\E}[1]{\mathrm{e}^{\mbox{\footnotesize$#1$}}}
\newcommand{\I}{\mathrm{i}}
\newcommand{\DET}[1]{\mathrm{Det}\{#1\}}
\newcommand{\PR}{\mathrm{pr}}
\newcommand{\erf}[1]{\mathrm{erf}\left(#1\right)}
\newcommand{\D}{\mathrm{d}}

\begin{document}
	
	\title{Experimental Evidence-Based Sub-Rayleigh Source Discrimination}  
	
	\author{Saurabh U. Shringarpure}
	\email{saurabh.s@snu.ac.kr}
	\affiliation{NextQuantum Innovation Research Center, Department of Physics and Astronomy,  Seoul National University, Seoul 08826, South Korea}
	
	\author{Yong Siah Teo}
	\email{ys\_teo@snu.ac.kr}
	\affiliation{NextQuantum Innovation Research Center, Department of Physics and Astronomy,  Seoul National University, Seoul 08826, South Korea}
	
	\author{Hyunseok Jeong}
	\email{h.jeong37@gmail.com}
	\affiliation{NextQuantum Innovation Research Center, Department of Physics and Astronomy,  Seoul National University, Seoul 08826, South Korea}
	
	\author{Michael Evans}
	\email{mevansthree.evans@utoronto.ca}
	\affiliation{Department of Statistical Sciences, University of Toronto, Toronto, Ontario, M5S 3G3, Canada}
	
	\author{Luis~L. S\'anchez-Soto}
	\affiliation{Departamento de \'Optica, Facultad de F\'{\i}sica, Universidad Complutense, 28040~Madrid, Spain}
	\affiliation{Institute for Quantum Studies, Chapman University, Orange, CA 92866, USA}
	\affiliation{Max-Planck-Institut f\"ur die Physik des Lichts, 91058 Erlangen, Germany}
	
	\author{Antonin Grateau}
	\affiliation{Laboratoire Kastler Brossel, Sorbonne Universit\'e, ENS-Universit\'e PSL, CNRS, Coll\`ege de France, 4 place Jussieu, F-75252 Paris, France} 
	
	\author{Alexander Boeschoten}
	\affiliation{Laboratoire Kastler Brossel, Sorbonne Universit\'e, ENS-Universit\'e PSL, CNRS, Coll\`ege de France, 4 place Jussieu, F-75252 Paris, France} 
	
	\author{Nicolas Treps}
	\affiliation{Laboratoire Kastler Brossel, Sorbonne Universit\'e, ENS-Universit\'e PSL, CNRS, Coll\`ege de France, 4 place Jussieu, F-75252 Paris, France}

	\begin{abstract}
We propose a Bayesian evidence-based inference framework based on relative belief ratios and apply it to discriminating between one and two  incoherent optical point sources using  spatial-mode demultiplexing~(SPADE). Unlike the Helstrom measurement, SPADE  require no collective detection and its optimal for asymptotically large samples.  Our method avoids \emph{ad hoc} statistical constructs and relies solely on the information contained in the data, with all assumptions entering only through the likelihood model and prior beliefs.  Using experimental evidence, we demonstrate the superior resolving performance of SPADE over direct imaging from a new and extensible perspective; one that naturally generalizes to multiple sources and offers a practical robust approach to analyzing quantum-enhanced superresolution.
	\end{abstract}
	\date{\today} 
	\maketitle

	\emph{Introduction.---} Quantum hypothesis testing stands as a cornerstone of quantum information science~\cite{Helstrom:1976aa,Hayashi:2017aa,Wilde:2017aa}, addressing one of its most essential challenges: how to distinguish quantum states (or channels) with maximal precision. At its heart lies the quest to identify the ultimate quantum limits of such distinguishability.
	
	The foundational works of Helstrom~\cite{Helstrom:1969aa} and Holevo~\cite{Holevo:1973aa} laid the theoretical bedrock for this field, deriving optimal strategies and fundamental bounds for discriminating between two nonorthogonal quantum states. While the two-state problem is now well understood, generalizing these insights to more complex scenarios remains an open challenge, with optimal measurements known only in special cases~\cite{Ban:1997aa}. Over the years, various approaches have been developed to identify measurements tailored to different performance metrics~\cite{Barnett:2009aa}, ranging from minimum-error and maximum-confidence approaches~\cite{Croke:2006aa} to unambiguous state discrimination~\cite{Chefles:1998aa,Rudolph:2003aa,Ivanovic:1987aa,Dieks:1988aa,Peres:1988aa,bergou2010discrimination,Bergou:2012aa}.
	
	In optical imaging, a common hypothesis testing problem is source discrimination; i.e., determining whether an image originates from one or two pointlike sources~\cite{Harris:1964aa,Helstrom:1973aa,Cunningham:1976aa,Acuna:1997aa,Shahram:2006aa}. This task is especially relevant in advanced imaging contexts such as exoplanet detection~\cite{Tsang:2019aa,Huang:2021aa} and fluorescence microscopy~\cite{Lee:2012aa,Nan:2013aa,Backlund:2018aa}. Yet, when the separation between sources is smaller than the system point spread function~\cite{Goodman:2004aa}, the performance of direct imaging plummets~\cite{Dekker:1997aa}.
	
	In this sub-Rayleigh regime, quantum-inspired measurement strategies that exploit the optimality of modal projection  for parameter estimation ~\cite{delaubertTEM_10Homodyne2006} have emerged as powerful alternatives to conventional imaging.  Seminal work by Tsang and collaborators~\cite{Tsang:2016aa,Nair:2016aa,Ang:2016aa} on estimating the separation of incoherent sources triggered a broad line of research~\cite{Paur:2016aa,Yang:2016aa,Tham:2017aa,Rehacek:2017aa, Zanforlin:2022aa,Grace:2022aa,Schlichtholz:2024aa,Wadood:2024aa}, driven in large part by the development of spatial-mode demultiplexing (SPADE)~\cite{Labroille:2014aa}. Subsequent work also established the optimality of SPADE for source discrimination~\cite{Lu:2018aa}.
	
	However, these promising results often rely on idealized measurement conditions. In practice, SPADE  performance is vulnerable to various imperfections, such as misalignments and fabrication defects, that introduce modal crosstalk and degrade measurement fidelity~\cite{Nichols:2016aa,Grace:2020aa,Lupo:2020aa,Len:2020aa,oh:2021aa,Sorelli:2021aa,Almeida:2021aa,Gessner:2020aa,Sorelli:2021ab,Linowski:2023aa}.  
	
	In this Letter, we address these limitations by adopting a Bayesian, evidence-based approach to state discrimination of light sources. Recasting the problem as one of hypothesis testing, we employ relative belief (RB) analysis~\cite{Evans:2015aa,Evans:2016aa,Evans:2018aa}, originally introduced in quantum information processing to construct concise and optimal credible regions for quantum state estimation and incomplete tomography with limited data~\cite{Shang:2013aa,Li:2016aa,Teo:2024ab,Teo:2024aa, Agarwal:2025, Englert:2025}. RB analysis requires only three essential components:  a well-specified model,  the observed measurement data, and  prior distributions, with optional checks for prior bias.
	
	This approach avoids \emph{ad hoc}  statistical assumptions, such as significance levels and $p$-values  commonly invoked in conventional hypothesis testing, and  readily accommodates model-specific hypotheses, making it suitable for many applications.  While frequentist methods based on Fisher information~\cite{Kay:1993aa}  are valuable for assessing measurement and estimator optimality, they are generally valid only in the asymptotic limit of many repeated measurements.   In contrast, the RB framework remains reliable even with limited data, providing a meaningful quantification of statistical evidence without  large-sample approximations. This enables direct evaluation of hypothesis plausibility across data regimes. All assumptions enter transparently through the likelihood model and chosen priors, with no additional auxiliary assumptions required.
	
	\emph{Relative-belief analysis.---} Let us recall the basics of RB.  It provides a mathematical framework to quantify statistical evidence across a set of competing hypotheses,  rather than relying solely on posterior probabilities. Consider a set of $K$ hypotheses about a quantum system. Before conducting any experiment, we may assign prior probabilities $\PR(k)$ reflecting our initial beliefs or knowledge, possibly informed by earlier results.    After collecting data $\mathbb{D}$ from an experiment, these beliefs are updated to posterior probabilities $\PR(k|\mathbb{D})$,  representing our revised confidence in each hypothesis via Bayes’ rule: 
	\begin{equation}
		\PR(k|\mathbb{D})=\dfrac{L(\mathbb{D}|k)\PR(k)}{\sum^K_{k^{\prime}=1}L(\mathbb{D}|k^{\prime})\PR(k^{\prime})} \, .
	\end{equation}
	Here,  $L(\mathbb{D}|k)$ is the likelihood of observing the dataset $\mathbb{D}$ assuming that hypothesis~$k$ is true. Then RB directly compares the prior and posterior beliefs via 
	\begin{equation}
		\RB_k =\dfrac{\PR(k|\mathbb{D})}{\PR(k)} \, ,
		\label{eq:RBratio}
	\end{equation} 
	which measures whether our belief in hypothesis $k$  has increased due to the data. A value $\RB_k >1$ indicates that data supports hypothesis~$k$ and that our belief in it has~strengthened.
	
	However, the RB ratio greater than one does not convey how strongly we should believe in what the evidence indicates.  This is measured necessarily by a posterior probability. In the case of two possible states, this is measured by the posterior probability of the hypothesized state. When there is evidence in favor of the hypothesis, a high posterior probability indicates strong evidence in favor while, when there is evidence against, a small posterior indicates strong evidence against. When there are more than two possible states, then a comparison among the RB ratios is made and the strength is measured by the \emph{evidence strength}, naturally defined as
	\begin{equation}
		E_k={\sum_{k^{\prime}\neq k}}\PR(k^{\prime}|\mathbb{D})\Theta(\RB_{k}-\RB_{k^{\prime}}) \, ,
		\label{eq:evid-strength}
	\end{equation}
	where $\Theta(\cdot)$ is the Heaviside step function, here defined as zero for the null argument. 
	
	The RB framework introduces a clear and natural threshold at $\RB_k =1$, distinguishing hypotheses that are supported by the evidence from those that are not.  This contrasts sharply with traditional hypothesis-testing methods—such as reliance on $p$-values— that   are designed to reject a null hypothesis rather than quantify evidence for an alternative. Commonly used thresholds (e.g., $p=0.05$)  are arbitrary and not dictated by the data. A small $p$-value does not validate an alternative hypothesis; it merely allows rejection of the null under subjective criteria.  By contrast, RB offers a transparent and evidence-driven approach, emphasizing the strength of support the data provide for each hypothesis rather than how effectively they refute another~\cite{Benjamin:2018aa,Wasserstein:2016aa}.
	
	\begin{figure}[t]
		\centering
		\includegraphics[width=.90\columnwidth]{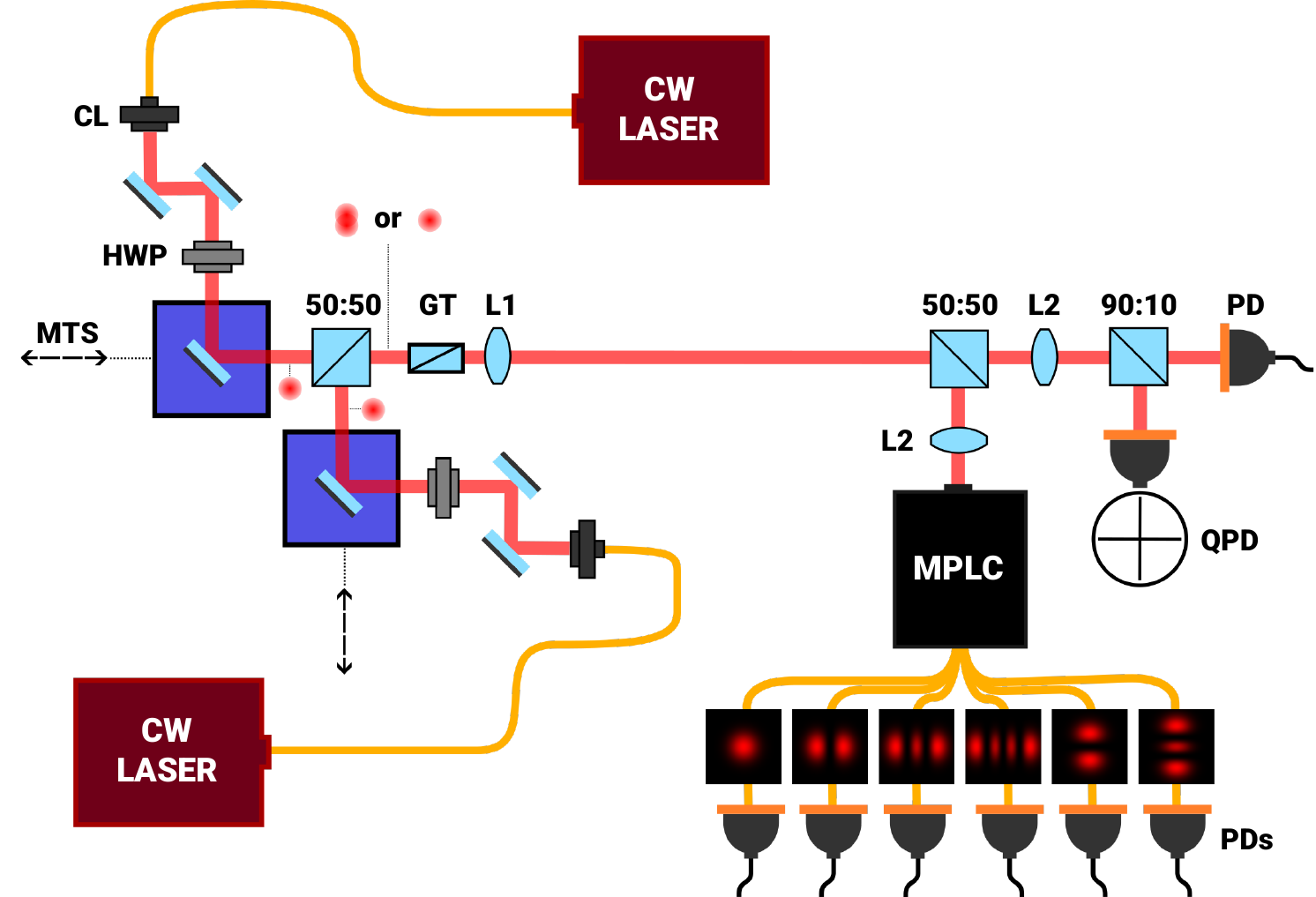}
		\caption{Schematic representation of the measurement setup. See the text for a detailed description.}
		\label{fig:meas}
	\end{figure}
	
	\emph{Measurement setting.---} Our setup, schematized in Fig.~\ref{fig:meas}, employs two independent, fiber-coupled CW lasers at 1550~nm to generate coherent states of light. Each output passes through a polarization controller and fiber attenuator for independent control of polarization and optical power. The resulting incoherent modes are collimated into free space as Gaussian beams with waist $w_{1} \simeq 1.135$~mm. Each beam is reflected by a mirror on a motorized translation stage, allowing fine adjustment of transverse displacement and precise control over spatial separation at the beam splitter. The combined beams are projected through a Glan--Thomson polarizer to ensure a common linear polarization, rendering the sources indistinguishable. A telescope images and mode-matches the beams for optimal coupling into a spatial-mode demultiplexer with waist $w_{0} \simeq 320~\mu$m.  
	
	For demultiplexing, we use a multiplane light converter (MPLC, PROTEUS-C from Cailabs) that decomposes the incoming light into the Hermite--Gauss (HG) mode basis. Each mode is subsequently coupled into a single-mode fiber. We utilize six of the ten available MPLC outputs and measure simultaneously their intensities using photodiodes. $\mathrm{HG}_{00}$, $\mathrm{HG}_{10}$, $\mathrm{HG}_{20}$, and $\mathrm{HG}_{30}$ are used for the test. In the preliminary stage dedicated to alignment and mode matching, we rely on modes $\mathrm{HG}_{10}$, $\mathrm{HG}_{01}$, $\mathrm{HG}_{20}$, and $\mathrm{HG}_{02}$. To accurately characterize the parameters of a given scene, we employ a quadrant detector along with an additional photodiode to sequentially measure the position and power of each source. In our protocol, we make no assumptions about the specific form of experimental imperfections, relying instead on robust model calibration.
	Any remaining subjective assumptions enter only through the explicit specification of the likelihood and the priors, rather than through \emph{ad hoc} thresholds or large-sample approximations. We also assess the robustness of our conclusions by varying the priors associated with nuisance parameters that may be sensitive to drift.
			
	\begin{figure}[t]
		\centering
		\includegraphics[width=.90\columnwidth]{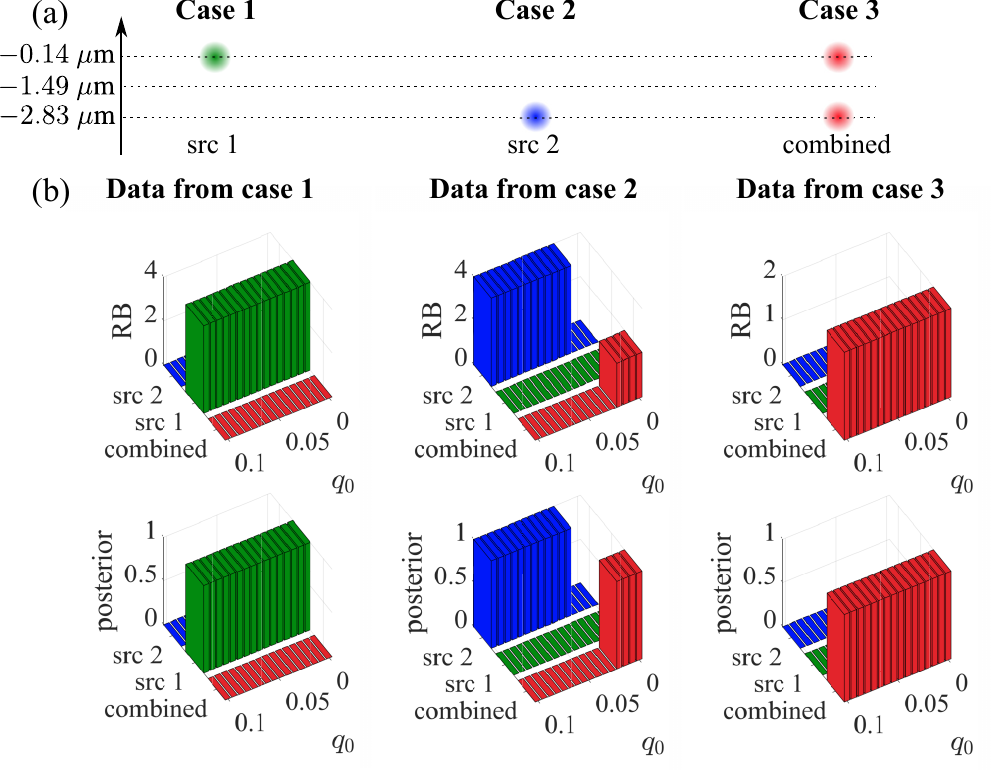}
		\caption{(a)~Two individual sources (src~1 and src~2), located at positions not aligned with the demultiplexer, illuminate the scene either individually or simultaneously (combined, with brightness ratio  $q=1/2$).  The source centroids $x_c$ for all three cases are marked.
		(b) RB-based source discrimination for data from all three illumination cases. For the two-source hypothesis, all centroid positions  $x_{c}$  within a sufficiently broad interval and all separations $d \neq 0$ are included, while brightness imbalance $q$ within a tolerance  $q_{0}$ around 0 or 1 is attributed to the corresponding single-source hypotheses, while the mutually exclusive interval is assigned to the two-source hypothesis. With priors of $1/2$  for the two-source hypothesis and  $1/4$ for each single-source hypothesis, exactly one hypothesis per panel yields	$\mathrm{RB} >1$, with posterior probability near unity. The Rayleigh limit for these sources is approximately ~$326.57~\mathrm{\mu m}$, indicating that the results exhibit superresolution. The results shown remain unchanged if the 10000 data samples are partitioned into sets of 50 and RB analysis is performed on each set independently, indicating extremely small variance in the results of the analysis. The results, however, depend on the choice of nuisance parameter space, which is arbitrary.} 
		\label{fig:1v2}
	\end{figure}
	
	To account for the setup imperfections and variations in the object profiles, the demultiplexed mode intensities are recorded as functions of the positions of the sources. These data are then calibrated using a cubic spline fit, which is essential for accurately characterizing experimental imperfections. The measured intensities are subject to random fluctuations and are recorded as voltage signals from the detectors, with a sampling rate of 10~kHz.  From these signals, the mean and the variance are extracted over a time interval of one second.   
	
	A typical hypothesis for the image is modeled as an incoherent mixture of the calibrated, mode-sorted intensities~$\tilde{\bm{I}}_{1}(x)$ and~$\tilde{\bm{I}}_{2}(x)$ of the two sources, where each $\tilde{\bm{I}}(x)=(\tilde{I}^{(0)}(x)\,\,\tilde{I}^{(1)}(x)\,\,\tilde{I}^{(2)}(x)\,\,\tilde{I}^{(3)}(x))^\top$ is a column of four calibrated  HG spatial-mode intensity distributions in~$x$. This is expressed as 
	\begin{equation}
		\tilde{\bm{I}}_d(x_\mathrm{c}) := q \tilde{\bm{I}}_{1} (x_\mathrm{c}-d/2)+(1-q)\tilde{\bm{I}}_{2}(x_\mathrm{c}+d/2) \, ,
		\label{eq:model}
	\end{equation} 
	where $x_\mathrm{c}$ is the centroid relative to the demultiplexer reference, $d$ is the separation between the sources and $q$ quantifies their brightness imbalance. For notational simplicity, the position  $x$  in the intensities will be omitted henceforth.
	
	The likelihood density is simply a product of the conditional probabilities for all modes,  $L_d=\prod_{j=0}^{3} \PR{(I^{(j)}|\tilde{I}^{(j)}_d)}$, where $I^{(j)}$s are the observed intensities in four HG spatial modes. The conditional probability densities are taken as Gaussians $\exp [ -\tfrac{1}{2}(\bm{I}-\tilde{\bm{I}}_d)^\top\cdot\bm{\Sigma}^{-1}\cdot(\bm{I}-\tilde{\bm{I}}_d)]\big/({2\pi \DET{\bm{\Sigma}}})$ with $\bm{\Sigma}={\rm{diag}}(\sigma_0^2,\sigma_1^2,\sigma_2^2,\sigma_{3}^2)$.  With these preliminary notations and ideas, let us now address two different tasks: source discrimination and parameter estimation. 
	
	\emph{Source discrimination.---}  First, we consider the source discrimination problem with three mutually exclusive hypotheses:  (i) a single source located at $x_{c} - d/2$, (ii) a single source located at $x_{c} + d/2$, or (iii) two sources present simultaneously at both positions,   with a brightness imbalance $q$ between them as specified in the model Eq.~\eqref{eq:model}.  Since $x_{c}$, $d$, and $q$ 	are unknown, they are treated as nuisance parameters. The three hypotheses are constructed by partitioning a sufficiently large and finely discretized nuisance-parameter space into mutually exclusive spaces corresponding to each hypothesis by an appropriate choice of priors. For the brightness imbalance $q$ they are uniform and nonzero in distinct intervals. For source separation, the $d=0$ point is assigned zero prior for the 2-sources hypothesis. The prior on centroid location is chosen to be uniform over a large interval for all three hypotheses. 
     The corresponding likelihoods are obtained by averaging over these priors (see section A in SM). 
     Evidence from RB analysis is then used to select one of the three hypotheses, in contrast to traditional reject-or-not hypothesis tests. The results in panel (b) of Fig.~\ref{fig:1v2} showcase the power of this approach: the data clearly indicate a single plausible hypothesis, reflected by a large RB value and a posterior probability close to unity (with a small tolerance $q_0$ allowed in $q$  to accommodate experimental variations). No additional \emph{ad hoc} assumptions are required to reach this conclusion.

	\begin{figure}[t]
		\centering
		\includegraphics[width=\columnwidth]{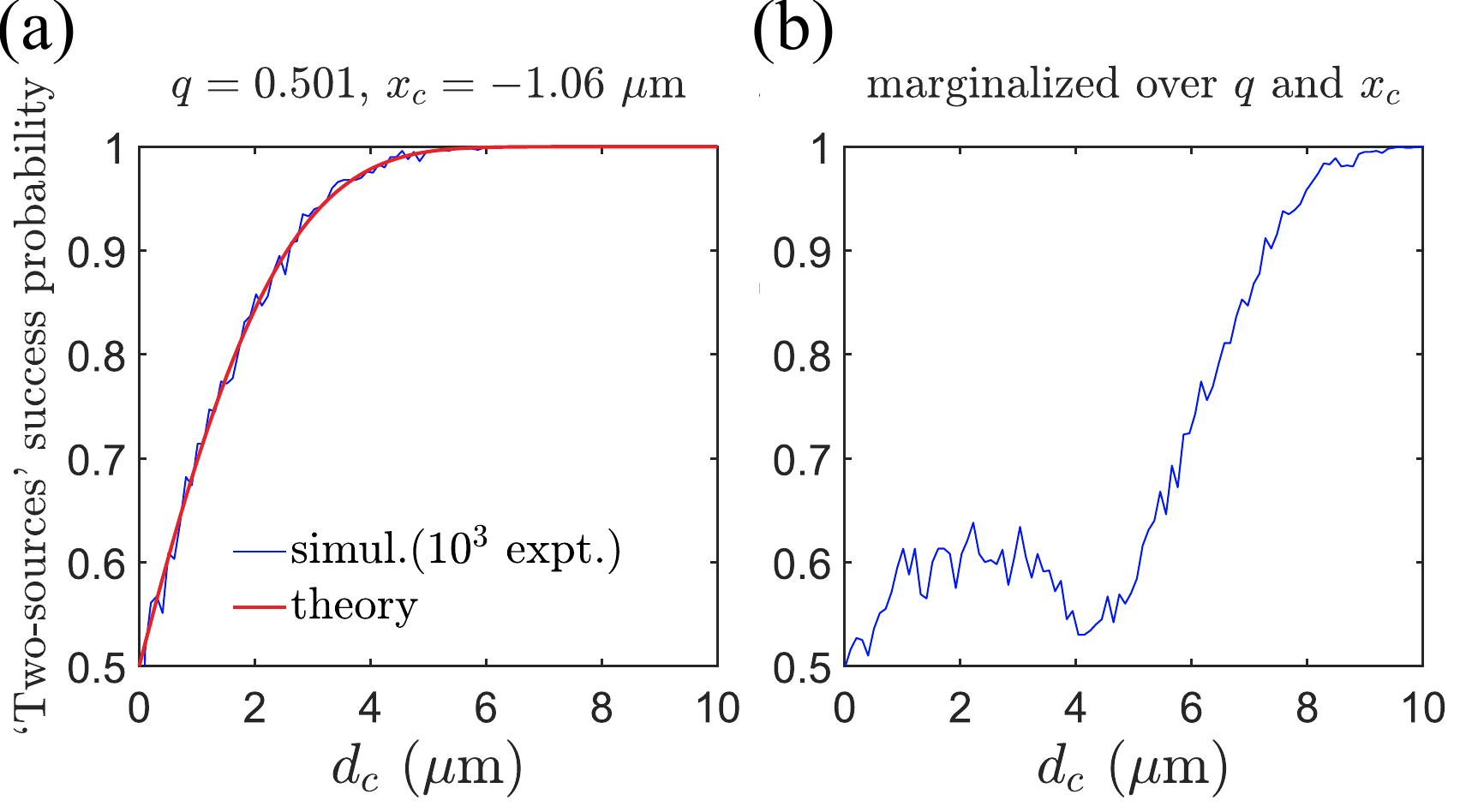}
		\caption{Sensitivity of RB in discriminating `one-source’ and `two-sources’ hypotheses.  Variances of the mode-sorted intensities from the experimental data with $x_{c} = -1.06~\mathrm{\mu m}$ and $q = 0.501$, together with models generated from the experimental calibration data, were used to simulate 1000 experiments of two sources separated by distance $d$. For each simulated experiment, the hypothesis with the maximum RB was selected.  The `two-sources' success probability is plotted against the source separation using simulated experiments (blue) for (a) known fixed $x_c$ and $q$, or (b) averaging over their unknown values. The expectation from theory is shown in red for the simpler problem in (a). In both cases, superresolution is achieved with $d_c \ll 326.57~\mathrm{\mu m}$ and error probability of $\epsilon < 10^{-3}$.}
		\label{fig:dc}
	\end{figure}
	
Now, let us understand the above experimental results in terms of the sensitivity of RB results to the source separation through a simpler problem with fixed centroid location and brightness imbalance, and demonstrate superresolution. Consider data generated by two sources separated by a distance $d$, then the RB for the `two-sources' hypothesis is given by
		\begin{equation}
			\RB_d = \frac{L_d}{\PR(1S) L_0 + \PR (2S) L_d} \, 
		\end{equation}
where we assume uniform priors $\PR{(1S)}=\PR{(2S)}={1}/{2}$ for complete ignorance.
	
Based on the evidence from the data, condition $\RB_d>1$ is used to decide the signal is from `two-sources', for which the success probability~reads
	\begin{equation}
		p_{\rm{succ}}= \tfrac{1}{2}\left[1+\erf{\sqrt{\bm{b}^\top \cdot\bm{\Sigma}^{-1}\cdot\bm{b}}/(2\sqrt{2})}\right] \, ,
		\label{eq:psucc}
	\end{equation}
where $\bm{b}:=\bm{I}_d-\bm{I}_0$ is the difference in columns of mode intensities for the `two-sources' and `one-source' hypotheses constructed using the calibration data and $\erf{x}$ is the error function (see section B in Supplemental Material).
	
	\begin{figure}[t]
		\centering
		\includegraphics[width=\columnwidth]{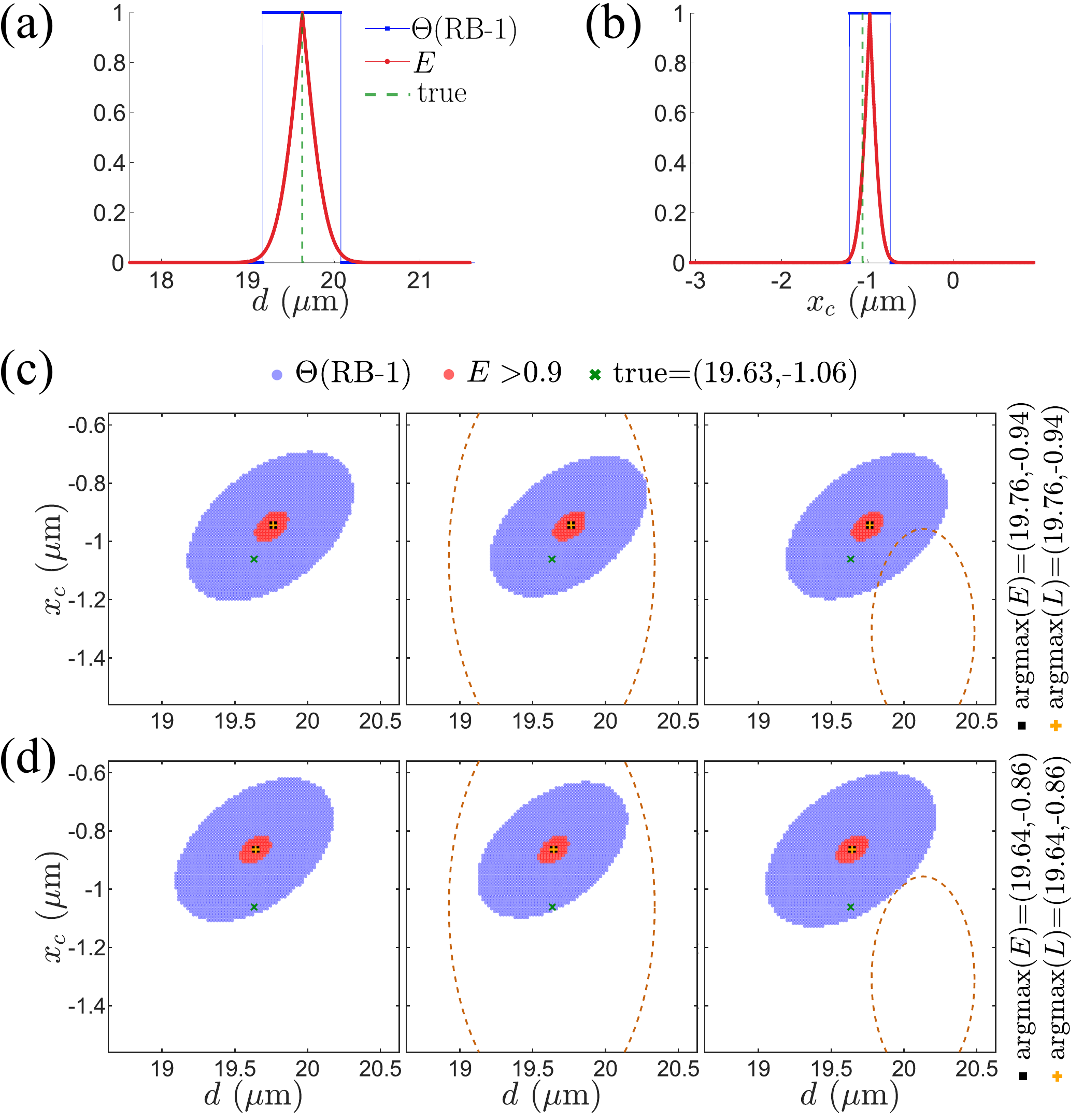}
		\caption{Parameter estimation using data-driven evidence from RB for equally bright  $(q=0.5)$ sources and varying either (a) the source separation $d$, (b) the centroid position~$x_\mathrm{c}$ in a $4~\mathrm{\mu m}$ interval with 5000 hypotheses using uniform priors, or (c,d) both in a $1~\mathrm{\mu m} \times 2~\mathrm{\mu m}$ region with $100 \times 100$ hypotheses around the parameter values determined for the experimental setup~(true), $d = 19.63~\mathrm{\mu m}$ and $x_\mathrm{c} = -1.06~\mathrm{\mu m}$ using uniform priors and Gaussian priors with variances of $0.5~\mathrm{\mu m}^2$ and $0.125~\mathrm{\mu m}^2$ centered at the true values and randomly elsewhere respectively where the dashed contours represent their standard deviation. All priors are normalized over the depicted space. All 10000 intensity readings were used in (c), whereas 10 were randomly selected in (d) to simulate a small sampling regime.}
	\label{fig:parameter}
\end{figure}

If the error probability, as given by Eq.~(\ref{eq:psucc}), satisfies \mbox{$1-p_{\rm{succ}} <\epsilon$} then  with Taylor expansions of the calibrated functions $\tilde{\bm{I}}_{1}(x_\mathrm{c}-d/2)$ and \mbox{$\tilde{\bm{I}}_{2}(x_\mathrm{c}+d/2)$} about $x_\mathrm{c}$, neglecting the $\mathcal{O}(d)$ term in $\bm{b}$ for nearly identical sources, and neglecting  $\mathcal{O}(d^3)$ terms for small $d$, yields a lower bound on the resolution
\begin{equation}
		d_\mathrm{c}\geq \left\{ \frac{8 [ \operatorname{erf}^{-1}(1-2\epsilon)]^{2}}{\bm{c}_2^\top \cdot\bm{\Sigma}^{-1}\cdot\bm{c}_2}\right\}^{\tfrac{1}{4}} \, ,
		\label{eq:d-critical}
	\end{equation}
where $\bm{c}_2:=[q \tilde{\bm{I}}^{\prime \prime}_{1}(x_\mathrm{c})+(1-q)\tilde{\bm{I}}^{\prime \prime}_{2}(x_\mathrm{c})]/8$ and $\tilde{\bm{I}}^{\prime \prime}_{1,2}(x)$ denotes the second derivatives. 

The data assessment using RB source discrimination  can serve as a  Bayesian postprocessing to extract the additional information revealed by the modal decomposition, thereby enhancing the resolved source separation \emph{without} requiring any optimization procedure, in contrast to maximum likelihood methods. This  surpasses the classical resolution limits imposed by the Rayleigh criterion~\cite{Ram:2006aa,Zhou:2019aa}, which states that two Gaussian objects can be resolved with direct imaging only when their separation exceeds the sum of their half-widths. The zeroth-order mode intensity from the calibration data corresponds to such Gaussian objects in the scene with $I^{(0)}\propto|\langle 0 | \E{-\I \sqrt{2} P x}|0\rangle|^2$ where $P$ is the momentum operator and the sources are displaced vacuua. Fitting $\tilde{I}^{(0)}_{1}$ and $\tilde{I}^{(0)}_{2}$ to Gaussians, $a \exp [-(x-b)^2/w^2]$, give the widths $w$  of $330.19~\mathrm{\mu m}$ and $322.94~\mathrm{\mu m}$ respectively, leading to a  Rayleigh limit of $326.57~\mathrm{\mu m}$.  In contrast, the critical distance of Eq.~\eqref{eq:d-critical} gives  $d_\mathrm{c} \simeq 6~\mathrm{\mu m}$ for an error probability of $10^{-3}$,  as can be seen in Fig.~\ref{fig:dc}. This demonstrates the superresolution of~SPADE without relying on the saturation of the quantum Cram{\'e}r--Rao bound~(QCRB) for unbiased estimators.

\emph{Parameter estimation.---}
The RB paradigm allows  flexible assessment of hypothesis plausibility.  For instance,  if the scene contains two sources,   we may estimate either the \emph{plausible interval} of the source separation~$d$ or that of the centroid position~$x_\mathrm{c}$ individually (single-parameter estimation), provided the other parameter is reasonably known.  If neither parameter can be reliably ascertained, RB can be used to find the \emph{plausible region} of both unknown parameters simultaneously (multiparameter estimation). This is achieved by scanning hypotheses over a finely divided interval (for a single parameter) or a region (for multiple parameters) centered around the true values chosen for the experiment.

Figure~\ref{fig:parameter} summarizes the versatility of RB in transparently revealing the plausible parameter regions just based on the measurement data at~hand.  Nearly unchanged and small sized plausible regions are due to the small uncertainties in the intensities measured by SPADE. This approach showcases the flexibility of the RB: it can be used to evaluate any number of model parameters~systematically.

\emph{Concluding remarks.---} Optimal state discrimination is a cornerstone of quantum information processing, with far-reaching implications for practical applications such as quantum imaging.  Superresolution techniques like SPADE have redefined the limits  for resolving closely-spaced sources, with potential impact on applications from astronomy to biological microscopy. Traditionally, when the problem is viewed as a parameter estimation, the analysis has relied on quantum Fisher information and the QCRB, particularly for estimating source separation. Alternatively, state discrimination has also been framed as hypothesis testing problems using statistical significance measures such as the  $p$-value. However, both the QCRB-based and conventional hypothesis testing approaches to statistical analysis have come under increasing scrutiny for relying on implicit and often \emph{ad hoc} assumptions in the analysis itself.

To address these limitations, we have introduced an evidence-based framework for statistical inference grounded in the RB paradigm. This approach quantifies support for a hypothesis or model parameter by comparing the posterior to the prior, providing evidence for a particular belief precisely when the data raise its plausibility. In addition to identifying supported hypotheses, RB naturally ranks competing alternatives through the strength of this evidence. In this way, RB treats both types of state-discrimination problem within a single, unified framework. Crucially, this framework remains valid even in regimes with limited data, whereas other popular methods for reaching the QCRB carry no relevant meaning in such pre-asymptotic data-sample~regimes. It extends seamlessly to both single and multiparameter estimation problems and offers a promising path toward analyzing more complex scenarios, including discrimination among multiple sources. These features position RB as a powerful and flexible statistical-inference tool in quantum imaging and~beyond.

\emph{Acknowledgements.---} We acknowledge discussions with A. Datta, M. Barbieri, and  M. Tsang at an early stage of this work. This publication is supported by the National Research Foundation of Korea (NRF) grants funded by the Korean government (MSIT) (Grant Nos. NRF-2023R1A2C1006115, RS-2023-00237959, RS-2024-00413957, RS-2024-00437191, RS-2024-00438415, RS-2025-02219034), the Institute of Information Communications Technology Planning Evaluation (IITP) grant funded by the Korea government (MSIT) (IITP-2025-RS-2020-II201606 and IITP-2025-RS-2024-00437191), the Institute of Applied Physics at Seoul National University, the Spanish Agencia Estatal  de Investigaci\'on (Grant PID2021-127781NB-I00). L. L. S. S. was supported in part by the grant NSF PHY-1748958 to the Kavli Institute for Theoretical Physics (KITP). This project has received funding from the European Defence Fund (EDF) under grant agreement ``101103417 EDF-2021-DIS-RDIS-ADEQUADE”,  funded by the European Union. Views and opinions expressed are however those of the author(s) only and do not necessarily reflect those of the European Union or the European Commission. Neither the European Union nor the granting authority can be held responsible for them.

%

\newpage
\onecolumngrid

\begin{center}{\bf Supplemental Material}\end{center}
\appendix

\section{Likelihoods for source discrimination problem}

Here we explain how to evaluate the relative beliefs of two `one-source' hypotheses and a single `two-sources' hypothesis of the source discrimination problem where the brightness imbalance $q$, the centroid location $x$, and the source separation $y$ are so-called \emph{nuisance parameters} of the likelihood model. To account for their arbitrary values we integrate over their likelihood densities weighted by priors as
\begin{equation}
	L_{2S}:=\int_{0}^{1}\hspace{-0.5em}\D q\int_{-\infty}^\infty\hspace{-1em}\D x\int_{0}^\infty\hspace{-1em}\D y\,\,\PR(q,x,y)\,L(\mathbb{D}|q,x,y),
	\label{eq:2S-likelihood}
\end{equation}
for the `two-sources' hypothesis.
In this manner, we do not assume any particular $(q,x,y)$ triplet when using RB analysis for the two-sources hypothesis. On the other hand, for the single-source hypotheses, the priors would be profiles of $\PR(q,x,y)$ at $q=0$~(or 1) and $y=0$ and hence
\begin{align}
	L_{1S}:=\int_{-\infty}^\infty\hspace{-1em}\D x\,&\frac{\PR(0~(\textrm{or}\,1),x,0)}{\int_{-\infty}^\infty\!\D x'\,\PR(0~(\textrm{or}\,1),x',0)}L(\mathbb{D}|q=0~(\textrm{or}\,1),x,y=0),
	\label{eq:1S-likelihood}
\end{align}
where $q=0$ (or 1) refers to the 1S being either src 1 (or src 2) as shown, for example, in Fig.~2 of the main text. 
If however we are ignorant about the specific kind of single source, we may consider the likelihood
\begin{equation}
	L_{1S}:=\int_{-\infty}^\infty\hspace{-1em}\D x\,\bigg[\frac{\PR(0,x,0)}{\mathcal{N}}L(\mathbb{D}|0,x,0)+\frac{\PR(1,x,0)}{\mathcal{N}}L(\mathbb{D}|1,x,0)\bigg],
	\label{eq:1S-likelihood-source-type-ignorance}
\end{equation}
where ${\PR(0,x,0)}/{\mathcal{N}}$ and ${\PR(1,x,0)}/{\mathcal{N}}$ are the priors assigned to the source types~(src 1 or 2 here) and where
\begin{equation}
	\mathcal{N}:=\int_{-\infty}^\infty\hspace{-1em}\D x'\,\big[\PR(0,x',0)+\PR(1,x',0)\big],
\end{equation}
is the normalization. For the 1S hypotheses, a reasonable assumption could be that our prior on $q$ is spread over small intervals near 0 or 1 to allow for small perturbations to experimental setup after calibration. Let the tolerances be $q_0$, then the likelihoods of the three cases become
\begin{align}
	L_{\text{src 1}}\,&=\int_{0}^{q_0}\hspace{-1em}\D q\,\int_{-\infty}^\infty\hspace{-1em}\D x\,\PR(q,x,0)\,L(\mathbb{D}|q,x,0)\nonumber\\
	L_{\text{src 2}}\,&=\int_{1-q_0}^{1}\hspace{-1em}\D q\,\int_{-\infty}^\infty\hspace{-1em}\D x\,\PR(q,x,0)\,L(\mathbb{D}|q,x,0)\nonumber\\
	L_{\text{combined}}\,&=\int_{q_0}^{1-q_0}\hspace{-1em}\D q\,\int_{-\infty}^\infty\hspace{-1em}\D x\,\int_{0}^\infty\hspace{-1em}\D y\,\PR(q,x,y)\,L(\mathbb{D}|q,x,y)
\end{align} 
In any case, the relative beliefs of the three hypotheses are
\begin{align}
	\RB_{\text{src 1}}&=\frac{L_{\text{src 1}}}{\PR(\text{src 1})L_{\text{src 1}}+\PR(\text{src 2})L_{\text{src 2}}+\PR(\text{combined})L_{\text{combined}}},\nonumber\\
	\RB_{\text{src 2}}&=\frac{L_{\text{src 1}}}{\PR(\text{src 1})L_{\text{src 1}}+\PR(\text{src 2})L_{\text{src 2}}+\PR(\text{combined})L_{\text{combined}}},\nonumber\\
	\RB_{\text{combined}}&=\frac{L_{\text{combined}}}{\PR(\text{src 1})L_{\text{src 1}}+\PR(\text{src 2})L_{\text{src 2}}+\PR(\text{combined})L_{\text{combined}}}.
\end{align}
where $\PR(\text{src 1})$, $\PR(\text{src 2})$, and $\PR(\text{combined})$ are our priors for the three hypotheses.

\section{Success probability of source discrimination}

This section analyzes a toy problem of source discrimination where two hypotheses either a single `one-source' and a single `two-source' alternative are tested. Moreover, the `two-sources' hypothesis refers to a particular source separation $d$. We are interested here in understanding the sensitivity of the RB analysis in correctly identifying this `two-sources' hypothesis.

The relative belief of two sources separated by a distance $d$ is 
\begin{align}
	\RB_d=\frac{L_d}{\PR(1S) L_0 + \PR (2S) L_d}
\end{align}
where $\PR(1S)$ and $\PR(2S)$ are the prior probabilities assigned to the `one-source' and `two-source' hypotheses, and
\begin{equation}
	L_d=\PR(\bm{I}|\tilde{\bm{I}}_d)=\prod_{j=0}^{M-1} \PR{(I^{(j)}|\tilde{I}^{(j)}_d)}
	\label{eq:lik}
\end{equation}
is the likelihood density of observing the set of intensities $\bm{I}$ in $M$ measured modes $j=0,1,2,\ldots, M-1$ and where $\tilde{\bm{I}}_d$ are the model intensities. In the experiment, the number of modes is $M=4$.

The model intensities are defined as the mixture
\begin{equation}
	\tilde{\bm{I}}_d:=q\tilde{\bm{I}}_1(x_c-d/2)+(1-q)\tilde{\bm{I}}_2(x_c+d/2)
\end{equation}
where $\tilde{\bm{I}}_1(x)$ and $\tilde{\bm{I}}_2(x)$ are the calibrated intensities as functions of position, x, for the two different sources of interest.

The conditional probability densities, $\PR(\bm{I}|\tilde{\bm{I}}_d)$ appearing in the likelihood density are taken to be Gaussian,
\begin{equation}
	\PR(\bm{I}|\tilde{\bm{I}}_d)=\exp [ -\tfrac{1}{2}(\bm{I}-\tilde{\bm{I}}_d)^\top\cdot\bm{\Sigma}^{-1}\cdot(\bm{I}-\tilde{\bm{I}}_d)]\big/({\sqrt{2\pi}^{M} \DET{\bm{\Sigma}}}),
	\label{eq:cond}
\end{equation}
with the covariance matrix $\bm{\Sigma}:=\operatorname{diag}(\sigma_0^2,\sigma_1^2,\ldots,\sigma_{M-1}^2)$.

The success probability of correctly identifying a two-source alternative using relative belief is given by
\begin{equation}
	p_{\rm{succ}}=\int(\D\bm {I})\,L_d\Theta(\RB_d-1),
	\label{eq:psucc}
\end{equation}
wherein we substitute the expression from above.

The prior probabilities $\PR(1S)=1-\delta$ and $\PR(2S)=1+\delta$ for $-1/2<\delta<1/2$ give relative beliefs
\begin{align}
	\RB_d=&\frac{2L_d}{L_0+L_d+2\delta(L_d-L_0)}\cdot
\end{align}
The condition $\RB_d>1$ simplifies as follows:
\begin{align}
	&2L_d>L_0+L_d+2\delta(L_d-L_0)
	\nonumber\\
	\implies &(1-2\delta)L_d > (1-2\delta)L_0\nonumber\\
	\implies &L_d>L_0,
\end{align}
which is independent of the priors.

Substituting in the likelihood density from Eq.~(\ref{eq:lik}) and the conditional probability densities from Eq.~(\ref{eq:cond}), we rewrite the condition as
\begin{align}
	\E{-\tfrac{1}{2}(\bm{I}-\tilde{\bm{I}}_d)^\top\cdot\bm{\Sigma}^{-1}\cdot(\bm{I}-\tilde{\bm{I}}_d)} >& \E{-\tfrac{1}{2}(\bm{I}-\tilde{\bm{I}}_0)^\top\cdot\bm{\Sigma}^{-1}\cdot(\bm{I}-\tilde{\bm{I}}_0)} \nonumber\\
	(\bm{I}-\tilde{\bm{I}}_d)^\top\cdot\bm{\Sigma}^{-1}\cdot(\bm{I}-\tilde{\bm{I}}_d) <& (\bm{I}-\tilde{\bm{I}}_0)^\top\cdot\bm{\Sigma}^{-1}\cdot(\bm{I}-\tilde{\bm{I}}_0) \nonumber\\
	\implies\bm{b}^\top\cdot\bm{\Sigma}^{-1}\cdot\bm{X} + &\frac{1}{2}{\bm{b}^\top\cdot\bm{\Sigma}^{-1}\cdot\bm{b}}>0,
\end{align}
where we defined $\bm{b}:=\tilde{\bm{I}}_d-\tilde{\bm{I}}_0$ and a new random variable, $\bm{X}:=\bm{I}-\tilde{\bm{I}}_d$.

The success probability from Eq.~(\ref{eq:psucc}) can therefore be rewritten as
\begin{align}
	p_{\mathrm{succ}}=\frac{1}{\sqrt{2\pi}^M\DET{\bm{\Sigma}}}\int(\D \bm{X})\,\E{-\frac{1}{2}\bm{X}^\top\cdot\bm{\Sigma}^{-1}\cdot\bm{X}}\Theta(\bm{b}^\top\cdot\bm{\Sigma}^{-1}\cdot\bm{X} + {\bm{b}^\top\cdot\bm{\Sigma}^{-1}\cdot\bm{b}}/{2}).
	\label{eq:psucc-v2}
\end{align}
Now, using the integral representation of the Heaviside step function,
\begin{equation} 
	\Theta(x)=\lim_{\epsilon\to 0^+}\int\frac{\D\tau}{2\pi\I} \frac{\E{\I x\tau }}{\tau-\I\epsilon},
	\label{eq:Theta}
\end{equation}
we obtain,
\begin{align}
	p_{\mathrm{succ}}=&\lim_{\epsilon\to 0^+}\int \frac{\D \tau}{2\pi\I (\tau-\I\epsilon)}\frac{1}{\sqrt{2\pi}^M \DET{\bm{\Sigma}}}\int(\D \bm{X})\,\E{-\frac{1}{2}\bm{X}^\top\cdot\bm{\Sigma}^{-1}\cdot\bm{X}+\I\tau(\bm{b}^\top\cdot\bm{\Sigma}^{-1}\cdot\bm{X}+\frac{1}{2}\bm{b}^\top\cdot\bm{\Sigma}^{-1}\cdot\bm{b})}.
	\label{eq:psucc-v3}
\end{align}

The usual, multivariable Gaussian integral identity
\begin{align}
	\frac{1}{\sqrt{2\pi}^M\DET{\bm{\Sigma}}}\int(\D \bm{x})\,\E{-\frac{1}{2}\bm{x}^\top\cdot\bm{\Sigma}^{-1}\cdot \bm{x}+\bm{a}^\top\cdot\bm{x}}= \E{\frac{1}{2}\bm{a}^\top\cdot\bm{\Sigma}\cdot\bm{a}},
	\label{eq:gauss}
\end{align}
gives,
\begin{align}
	p_{\mathrm{succ}}=&\lim_{\epsilon\to 0^+}\int \frac{\D \tau}{2\pi\I \tau-\I\epsilon}\E{-\frac{\tau^2}{2}\bm{b}^\top\cdot\bm{\Sigma}^{-1}\cdot\bm{b}+\frac{\I \tau}{2}\bm{b}^\top\cdot\bm{\Sigma}^{-1}\cdot\bm{b}}.
	\label{eq:psucc-v4}
\end{align}

The above equation can be further simplified by using the identity,
\begin{equation}
	\frac{1}{\I(\tau-\I\epsilon)}=\int_0^\infty\D y\,\E{-\I y(\tau-\I\epsilon)}
	\label{eq:useful-identity}
\end{equation}
for $\epsilon>0$ which leads to,
\begin{align}
	p_{\mathrm{succ}}=&\lim_{\epsilon\to 0^+}\int_0^\infty\D y\,\E{-\epsilon y}\int\frac{\D\tau}{2\pi}\,\E{-\frac{\tau^2}{2}\bm{b}^\top\cdot\bm{\Sigma}^{-1}\cdot\bm{b}-\I\tau(y-\bm{b}^\top\cdot\bm{\Sigma}\cdot\bm{b}/2)}\nonumber\\
	=&\frac{1}{\sqrt{2\pi\bm{b}^\top\cdot\bm{\Sigma}^{-1}\cdot\bm{b}}}\int_0^\infty\D y\,\E{-\frac{(y-\bm{b}^\top\cdot\bm{\Sigma}^{-1}\cdot\bm{b}/2)^2}{2\bm{b}^\top\cdot\bm{\Sigma}^{-1}\cdot\bm{b}}},
	\label{eq:psucc-v5}
\end{align}
where the Gaussian integral identity from Eq.~(\ref{eq:gauss}) was used again. 

Now using the definite integral
\begin{equation}
	\frac{1}{\sqrt{2\pi}\sigma}\int_a^b\D x\,\E{-\frac{(x-\mu)^2}{2\sigma^2}}=\frac{1}{2}\erf{\frac{b-\mu}{\sqrt{2}\sigma}}-\frac{1}{2}\erf{\frac{a-\mu}{\sqrt{2}\sigma}},
\end{equation}
we finally get
\begin{equation}
	p_{\mathrm{succ}}=\frac{1}{2}+\frac{1}{2}\erf{\frac{1}{2}\sqrt{\frac{\bm{b}^\top\cdot\bm{\Sigma}^{-1}\cdot\bm{b}}{2}}}
	\label{eq:psucc-fin}
\end{equation}

\section{Critical distance of source discrimination of nearly indistinguishable Gaussian objects}

Suppose we upper bound some tolerance $\epsilon>0$ on the failure probability. Then
\begin{align}
	&1-p_{\mathrm{succ}}\leq\epsilon\nonumber\\
	\implies&\frac{1}{2}-\frac{1}{2}\erf{\frac{1}{2}\sqrt{\frac{\bm{b}^\top\cdot\bm{\Sigma}^{-1}\cdot\bm{b}}{2}}}\leq\epsilon\nonumber\\
	\implies&\bm{b}^\top\cdot\bm{\Sigma}^{-1}\cdot\bm{b}\leq 8[\operatorname{erf}^{-1}(1-2\epsilon)]^2.
	\label{eq:b-epsilon-condition}
\end{align}

Recall that 
\begin{align}
	\bm{b}=&\tilde{\bm{I}}_d-\tilde{\bm{I}}_0\nonumber\\
	=&q\left[\tilde{\bm{I}}_1(x_c-d/2)-\tilde{\bm{I}}_1(x_c)\right]+(1-q)\left[\tilde{\bm{I}}_2(x_c+d/2)-\tilde{\bm{I}}_2(x_c)\right].
\end{align}
Taylor's series of the calibration curves $\tilde{\bm{I}}_1(x_c-d/2)$ and $\tilde{\bm{I}}_2(x_c+d/2)$ about the point $x_c$ in the gives
\begin{align}
	\bm{b}=&\sum_{n=1}^\infty \frac{1}{n!}\left[q\left(-\frac{d}{2}\right)^n\frac{\D^n}{\D x^n}\tilde{\bm{I}}_1(x_c)+(1-q)\left(\frac{d}{2}\right)^2\frac{\D^n}{\D x^n}\tilde{\bm{I}}_2(x_c)\right],\nonumber\\
	\cong&d \bm{c}_1+d^2 \bm{c}_2 +\mathcal{O}(d^3)
	\label{eq:b}
\end{align}
where $\bm{c}_1:=\left[-q\tilde{\bm{I}}'_1(x_c)+(1-q)\tilde{\bm{I}}'_2(x_c)\right]/2$ and $\bm{c}_2:=\left[q\tilde{\bm{I}}''_1(x_c)+(1-q)\tilde{\bm{I}}''_2(x_c)\right]/8$. 

We neglect $\mathcal{O}(d^3)$ terms for further analysis, assuming that small $d$ and thus Eq.~(\ref{eq:b-epsilon-condition}) becomes
\begin{align}
	d^2\bm{c}_1^\top\cdot\bm{\Sigma}^{-1}\cdot\bm{c}_1+2 d^3 \bm{c}_1^\top\cdot\bm{\Sigma}^{-1}\cdot\bm{c}_2 +d^4 \bm{c}_2^\top\cdot\bm{\Sigma}^{-1}\cdot\bm{c}_2 \geq 8[\operatorname{erf}^{-1}(1-2\epsilon)]^2
	\label{eq:d-order-2-conditon}
\end{align}

Note that if the calibration for the two sources have identical profiles, and if the brightness imbalance, $q=1/2$, then $c_1\to 0$, in which case we would have
\begin{equation}
	d_c\geq \left\{\frac{8[\operatorname{erf}^{-1}(1-2\epsilon)]^2}{\bm{c}_2^\top\cdot\bm{\Sigma}^{-1}\cdot\bm{c}_2}\right\}^{\frac{1}{4}}.
\end{equation}
This is a conservative lower bound on the distance when the sources are nearly indistinguishable. 

\section{Effects of incorrect modeling}

In this section, we compile various results from incorrect data modeling.

\begin{figure}[htbp]
	\centering
	\includegraphics[width=0.5\columnwidth]{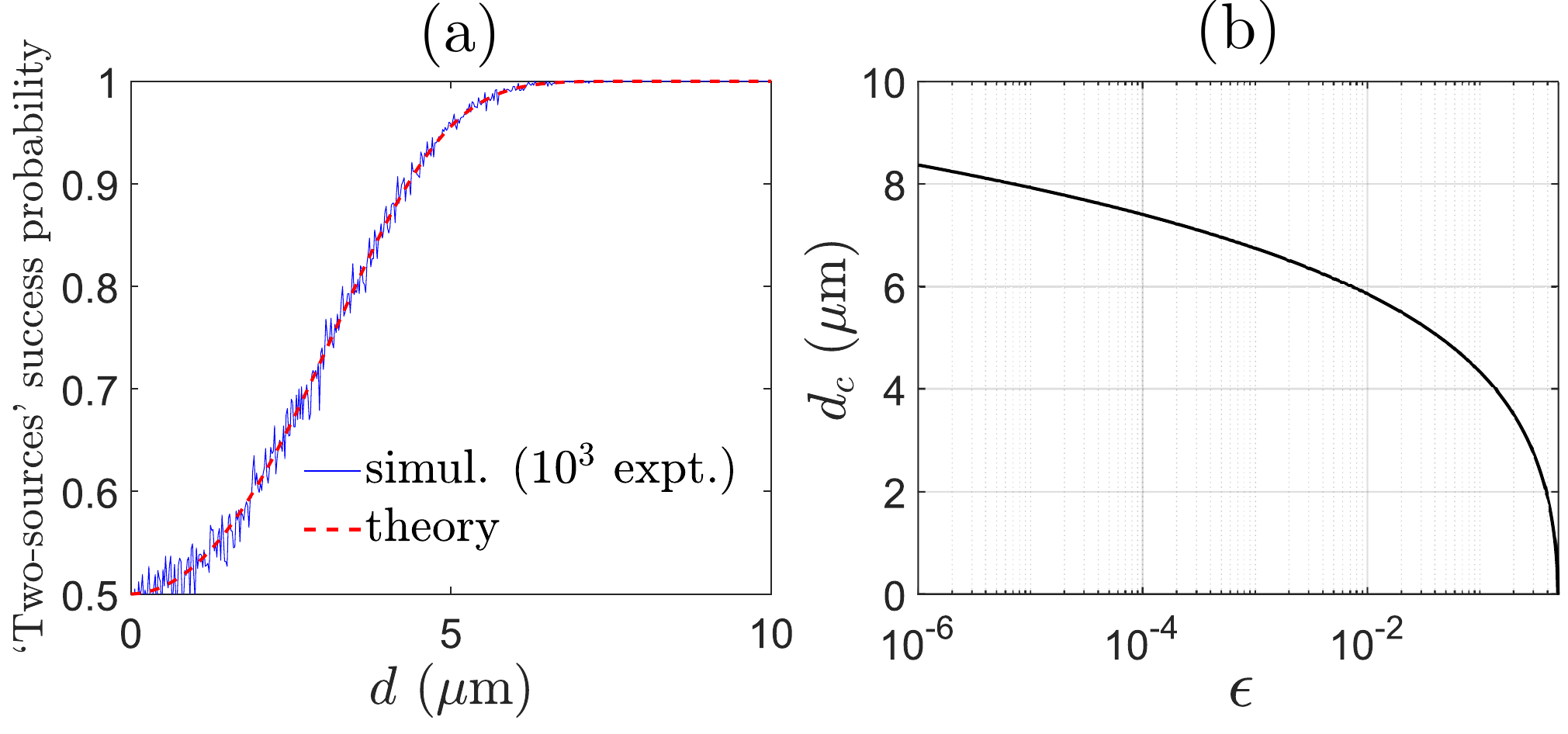}
	\caption{Sensitivity of RB in discriminating `one-source’ and `two-sources’ hypotheses. The parameters are identical to those in Fig. 3 of the main text. The simulated data and the theoretical model use calibration for only one source and assume the sources to be truly indistinguishable. As they match, we also see good agreement in the success probability of identifying the `two-sources' hypothesis, and the critical distance criterion can be used.}
	\label{fig:truly single}
\end{figure}
\begin{figure}[p]
	\centering
	\includegraphics[width=0.5\columnwidth]{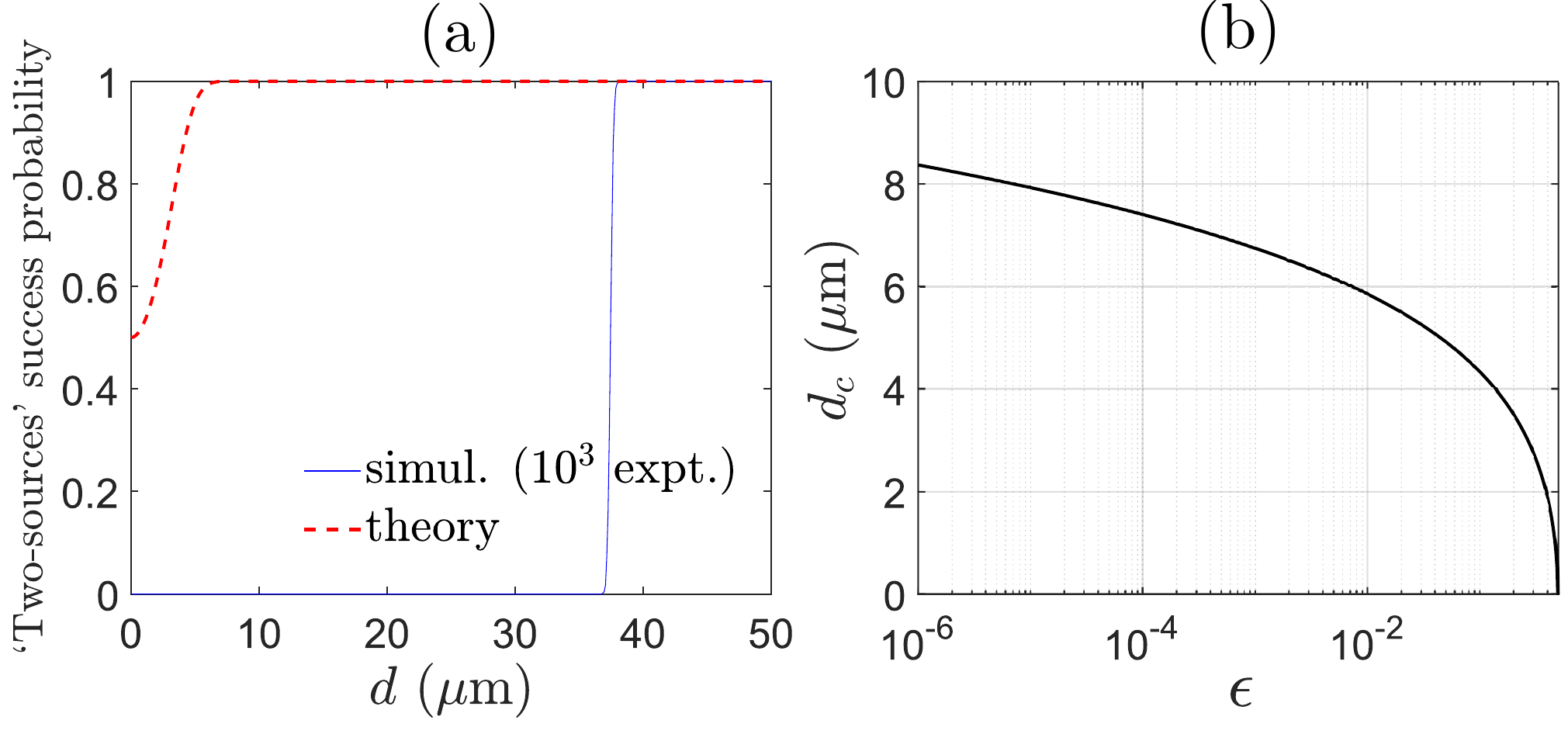}
	\caption{Sensitivity of RB in discriminating `one-source’ and `two-sources’ hypotheses using incorrect modelling. The parameters are identical to those in Fig. 3 of the main text. The simulated data uses calibration for both sources, assuming distinguishability, whereas the theoretical model uses calibration for only one source and assumes the sources to be truly indistinguishable. As the model does not represent the data, we observe a large discrepancy between simulation and theory, and the critical distance criterion is invalid. However, RB-based source discrimination becomes successful beyond a larger separation $\cong 37 \mu$m, which is still sub-Rayleigh distance $\cong 320\mu$m.}
	\label{fig:faulty single}
\end{figure}

\begin{figure}[bp]
	\centering
	\includegraphics[width=0.5\columnwidth]{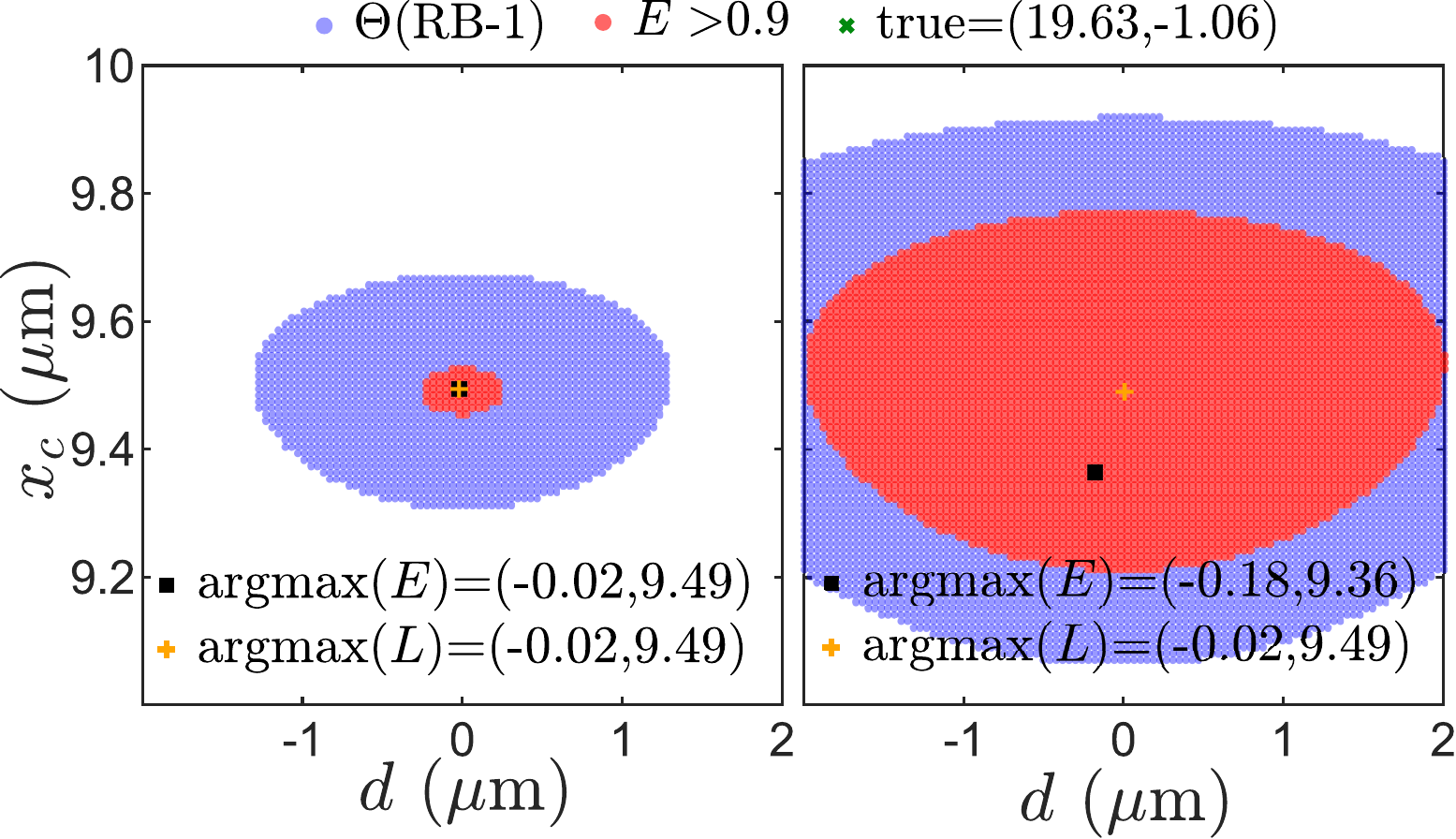}
	\caption{Plausible regions of parameters for the experimental data of Fig.~4(c) of the main text obtained by incorrectly modeling both sources with the calibration from a single source~(src 2 alone). Uniform priors and Gaussian priors with a variance of $0.5~\mathrm{\mu m}^2$ centered at the true value and normalized over the rectangular patch of parameters shown. All 10000 samples of the experimental data recorded over the acquisition time were used, but as the model clearly does not represent the data, we observe a large discrepancy and a greater extent of the plausible regions than in Fig.~4(c). Note that maximum likelihood is also susceptible to the incorrect choice of the model.}
	\label{fig:faulty single region}
\end{figure}

\end{document}